\documentclass{ws-procs9x6}

\usepackage{epsfig}
\newcommand{\agt}{\,\rlap{\lower 3.5 pt \hbox{$\mathchar \sim$}} \raise 1pt
 \hbox {$>$}\,}
\newcommand{\alt}{\,\rlap{\lower 3.5 pt \hbox{$\mathchar \sim$}} \raise 1pt
 \hbox {$<$}\,}

\begin{document}

\title{$D$-meson production in the GM-VFN scheme}

\author{B. A. Kniehl}

\address{II.~Institut f\"ur Theoretische Physik, Universit\"at Hamburg,\\
Luruper Chaussee 149, 22761 Hamburg, Germany\\
E-mail: kniehl@desy.de}

\maketitle

\abstracts{
We study the inclusive hadrodroduction of $D^0$, $D^+$, $D^{*+}$, and $D_s^+$ 
mesons at next-to-leading order in the parton model of quantum chromodynamics
endowed with universal non-perturbative fragmentation functions (FFs) fitted
to $e^+e^-$ annihilation data from CERN LEP1.
Working in the general-mass variable-flavor-number scheme, we resum the
large logarithms through the evolution of the FFs and, at the same time,
retain the full dependence on the charm-quark mass without additional
theoretical assumptions.
In this way, the cross section distributions in transverse momentum recently
measured by the CDF Collaboration in run~II at the Fermilab Tevatron are
described within errors.}

\section{Introduction}
\label{sec:one}

Recently, there has been much interest in the study of charmed-hadron ($X_c$)
production at hadron colliders, both experimentally and theoretically.
The CDF Collaboration measured the differential cross sections $d\sigma/dp_T$
for the inclusive production of $D^0$, $D^+$, $D^{*+}$, and $D_s^+$ mesons
(and their antiparticles) in $p\bar{p}$ collisions at the Fermilab Tevatron
(run II) as functions of transverse momentum ($p_T$) in the central rapidity
($y$) region.\cite{CDF}
Until recently, the most advanced theoretical predictions,\cite{BK,CN} based
on quantum chromodynamics (QCD) at next-to-leading order (NLO), consistently
undershot all the $D^0$, $D^+$, and $D^{*+}$ data by significant amounts, as
is evident from Fig.~3 of Ref.~\refcite{CDF}, while no predictions for $D_s^+$
mesons existed.
Especially in view of future physics at the CERN Large Hadron Collider, where
the continuum production of charmed hadrons will provide important backgrounds
for numerous new-physics signals, it is an urgent task to deepen our
understanding of the inclusive hadroproduction of charmed hadrons on the basis
of QCD in order to render the theoretical predictions as reliable as possible,
so as to establish a sturdy anchor for new-physics searches.
Here, we report on recent progress in this direction.\cite{Kniehl:2005ej}

\section{General-Mass Variable-Flavor-Number Scheme}

We wish to advocate the general-mass variable-flavor-number (GM-VFN) scheme,
which has recently been elaborated for the photoproduction\cite{KS} and
hadroproduction\cite{Kniehl:2004fy} of heavy-flavored hadrons.  
In this approach, one starts from the $p_T\gg m_c$ region and absorbs the large
logarithms $\ln(p_T^2/m_c^2)$ into the parton density function (PDF) of the
$c$-quark in the incoming hadrons and the fragmentation functions (FFs) for
the $c\to X_c$ transitions.
After factorizing the $\ln m_c^2$ terms, the cross section is infrared safe in
the limit $m_c\to0$, and $n_f=4$ is taken in the strong-coupling constant
$\alpha_s$ and the DGLAP evolution equations.  
The remaining $m_c$ dependence, i.e.\ the $m_c^2/p_T^2$ power terms, is
retained in the hard-scattering cross sections.
These terms are important in the intermediate $p_T$ region, where
$p_T\agt m_c$, and are expected to improve the precision of the theoretical
predictions.
The large logarithms are absorbed into the PDFs and FFs by subtraction of the
collinearly (mass) singular terms. 
However, in order to define a unique factorization prescription, one also has
to specify non-singular terms.
This is done by requiring that, in the limit $p_T\to\infty$, the known
hard-scattering cross sections of the zero-mass variable-flavor-number (ZM-VFN)
scheme are recovered. 
To achieve this, subtraction terms are derived by comparing the
fixed-flavor-number (FFN) theory in the limit $m_c\to0$ with the ZM-VFN theory,
implemented in the $\overline{\rm MS}$ factorization
scheme.\cite{KS,Kniehl:2004fy}
This matching procedure is useful, since all commonly used $c$-quark PDFs and
FFs are defined in the ZM-VFN scheme. 
The latter can then be used consistently together with hard-scattering cross
sections calculated in the GM-VFN scheme. 

\section{Numerical Results}

We are now in a position to present our numerical results for the cross
sections of inclusive $D^0$, $D^+$, $D^{*+}$, and $D_s^+$ hadroproduction to
be directly compared with the CDF data,\cite{CDF} which come as distributions
$d\sigma/dp_T$ at c.m.\ energy $\sqrt{S}=1.96$~TeV with $y$ integrated over
the range $|y|\le1$.
For each $X_c$ species, the particle and antiparticle contributions are
averaged.
We work in the GM-VFN scheme with $n_f=4$, thus excluding $X_c$ hadrons from
$X_b$-hadron decays, which are vetoed in the CDF analysis.\cite{CDF}
We set $m_c=1.5$~GeV and evaluate $\alpha_s^{(n_f)}(\mu_R)$, where $\mu_R$ is
the renormalization scale, with
$\Lambda^{(4)}_{\overline{\rm MS}}=328$~MeV,\cite{CTEQ6M} corresponding to
$\alpha^{(5)}(m_Z)=0.1181$.
We employ proton PDF set CTEQ6.1M from the CTEQ Collaboration\cite{CTEQ6M}
and the NLO FFs\cite{Kniehl:2005de} that were recently fitted to LEP1 data
taking the starting scales for the DGLAP evolution to be $\mu_0=m_c,m_b$.
We distinguish between the initial- and final-state factorization scales,
$\mu_F$ and $\mu_F^\prime$, so that we have three unphysical mass scales
altogether.
Our default choice is $\mu_R=\mu_F=\mu_F^\prime=m_T$, where
$m_T=\sqrt{p_T^2+m_c^2}$ is the transverse mass.
In order to conservatively estimate the theoretical error due to the scale
uncertainty, we independently vary the values of $\mu_R/m_T$, $\mu_F/m_T$, and
$\mu_F^\prime/m_T$ between 1/2 and 2, and determine the maximum upward and
downward deviations from our default predictions.

\begin{figure}[ht]
\begin{center}
\begin{tabular}{cc}
\parbox{0.5\textwidth}{\epsfig{file=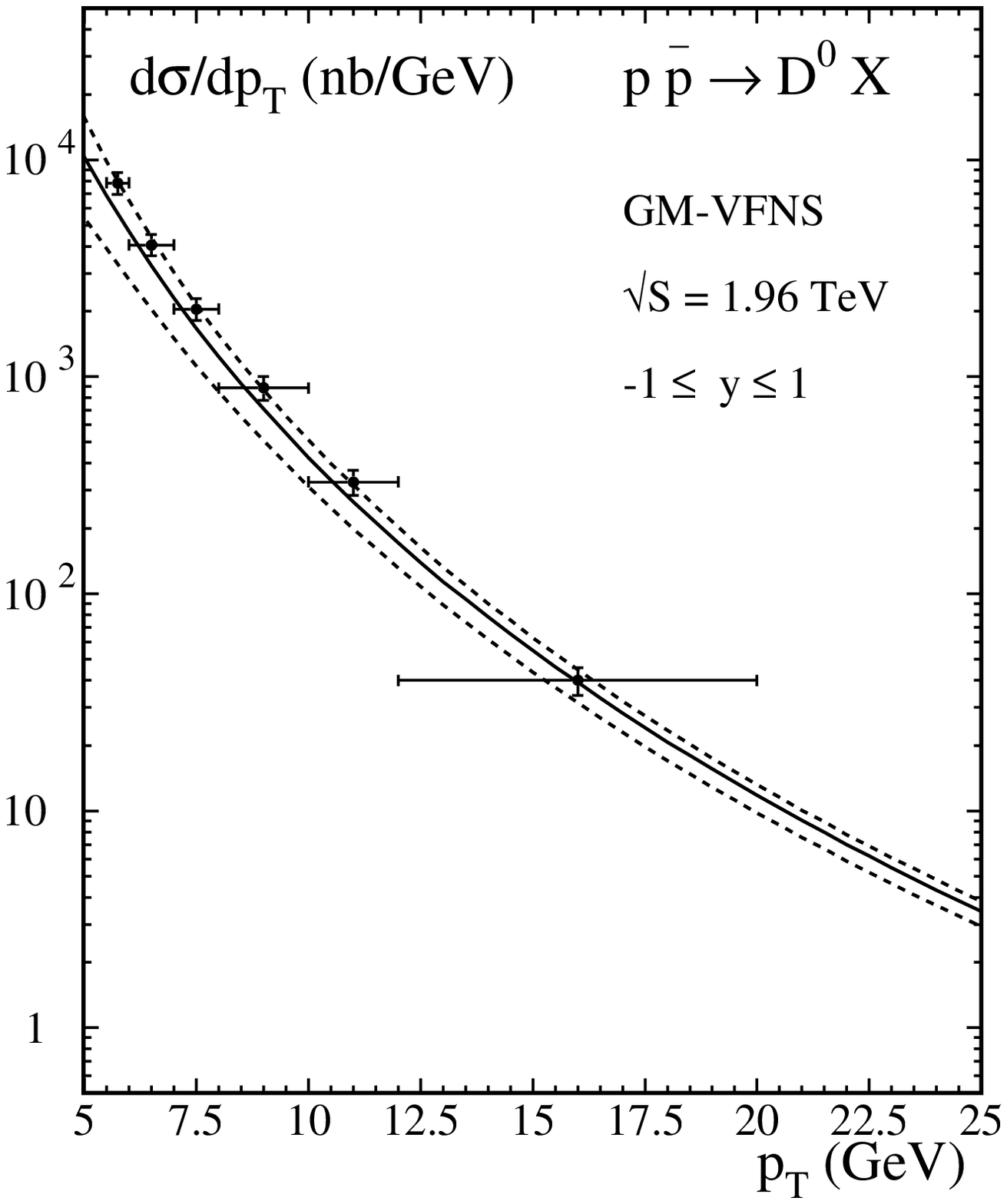,width=0.5\textwidth,%
bbllx=40pt,bblly=15pt,bburx=420pt,bbury=470pt}} &
\parbox{0.5\textwidth}{\epsfig{file=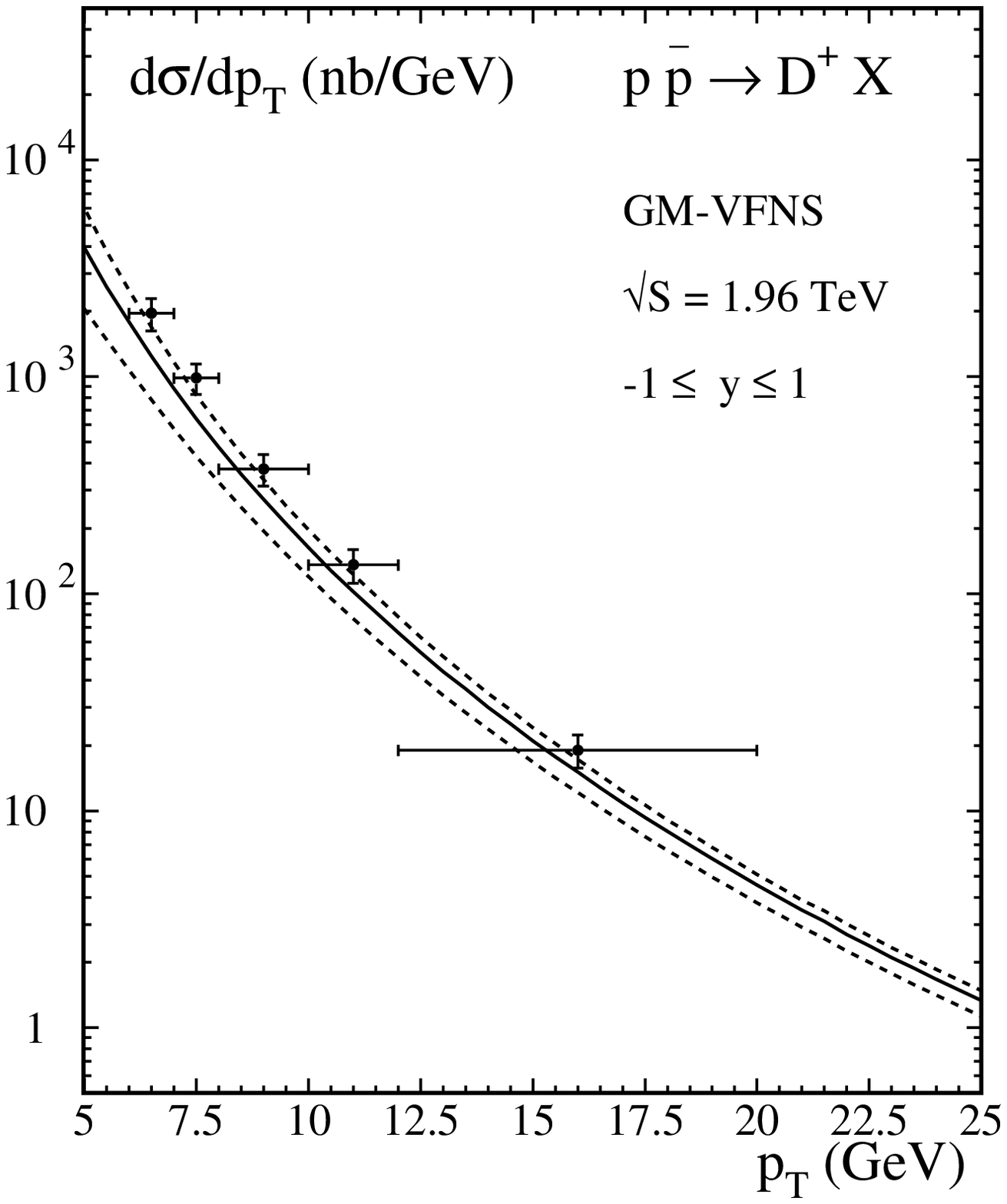,width=0.5\textwidth,%
bbllx=40pt,bblly=15pt,bburx=420pt,bbury=470pt}} \\
\parbox{0.5\textwidth}{\epsfig{file=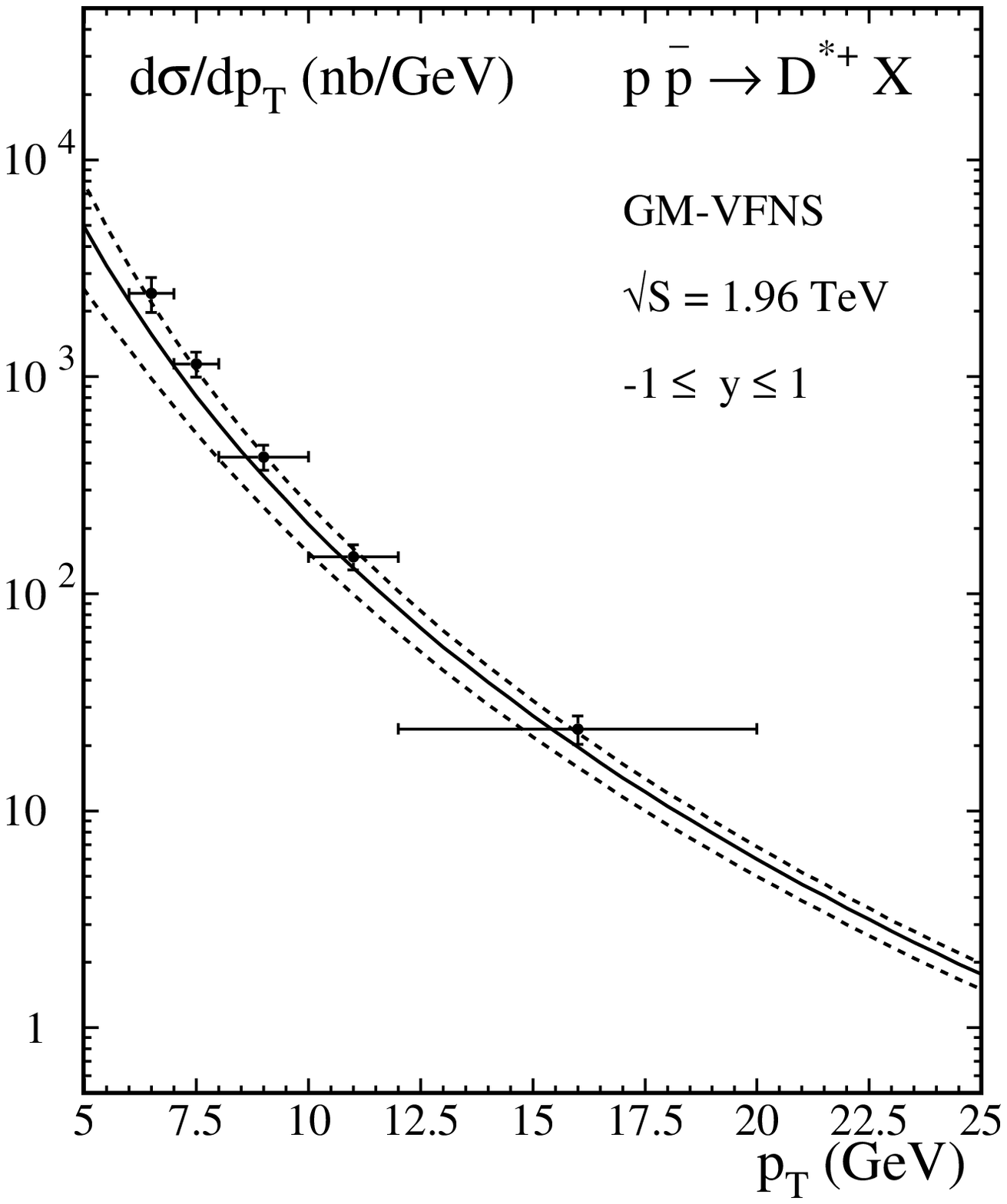,width=0.5\textwidth,%
bbllx=40pt,bblly=15pt,bburx=420pt,bbury=470pt}} &
\parbox{0.5\textwidth}{\epsfig{file=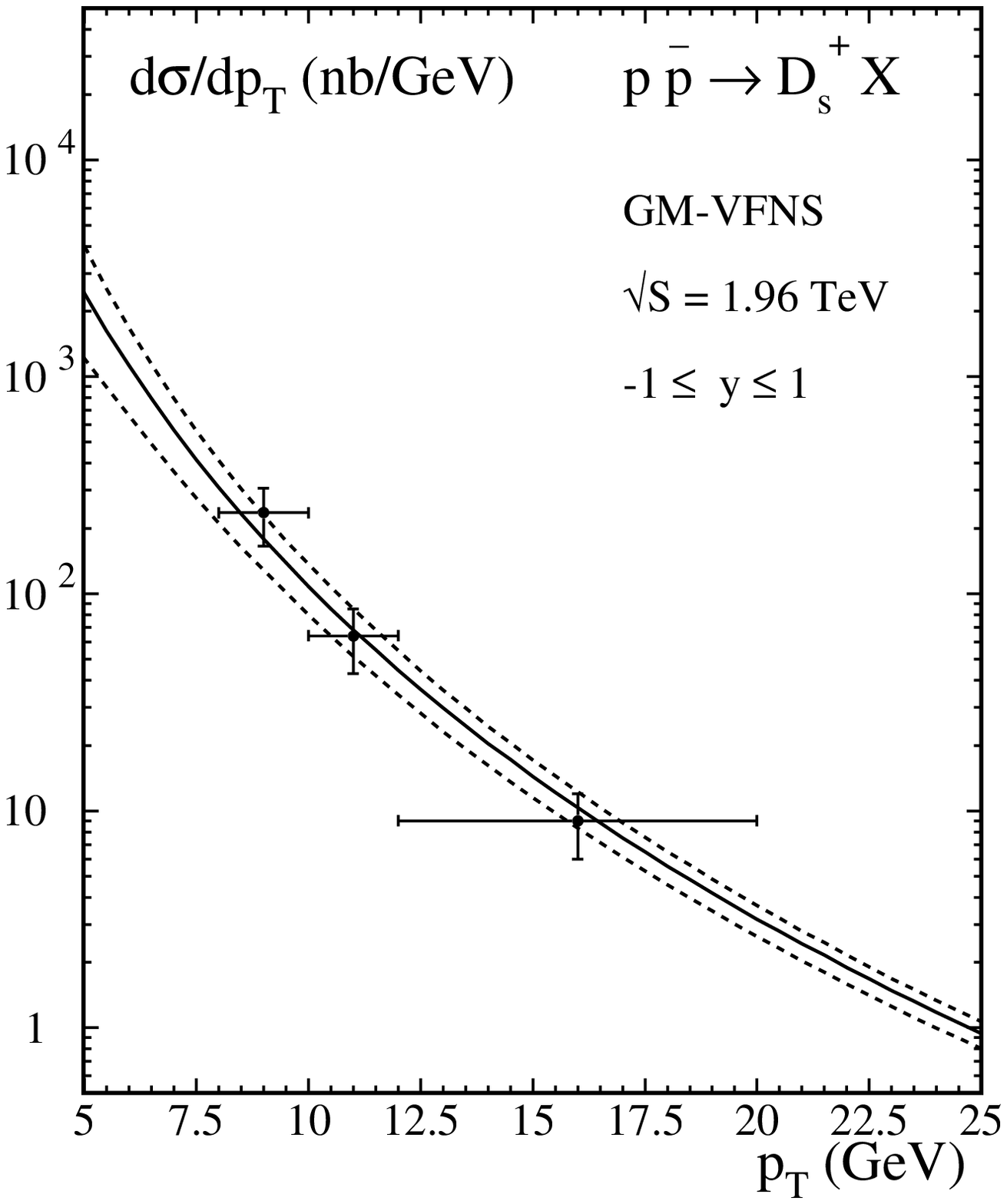,width=0.5\textwidth,%
bbllx=40pt,bblly=15pt,bburx=420pt,bbury=470pt}}
\end{tabular}
\end{center}
\caption{Comparison of the CDF data\protect\cite{CDF} with our NLO predictions
for $X_c=D^0,D^+,D^{*+},D_s^+$.
The solid lines represent our default predictions, while the dashed lines
indicate the unphysical-scale uncertainty.}
\label{fig:fig1}
\end{figure}

Our theoretical predictions are compared with the CDF data in
Fig.~\ref{fig:fig1}.
The four frames refer to $D^0$, $D^+$, $D^{*+}$, and $D_s^+$
mesons.
In all cases, we find good agreement in the sense that the theoretical and
experimental errors overlap, i.e.\ the notorious discrepancy between
experiment and theory mentioned in Sec.~\ref{sec:one} has disappeared.
In fact, our theoretical predictions provide the best description of the
CDF data obtained so far.

\section{Conclusions}

In conclusion, the GM-VFN scheme resums large logarithms by the DGLAP
evolution of non-perturbative FFs, guarantees the universality of the
latter as in the ZM-VFN scheme, and simultaneously retains the $m_c$-dependent
terms of the FFN scheme without additional theoretical assumptions.
Adopting this framework in combination with new fits of $D^0$, $D^+$,
$D^{*+}$, and $D_s^+$ FFs\cite{Kniehl:2005de} to LEP1 data, we managed for the
first time to reconcile the CDF data on the production of these mesons in
Tevatron run~II\cite{CDF} with QCD within errors and thus eliminated a
worrisome discrepancy.
Furthermore, we presented the first NLO predictions for the $D_s^+$
data.\cite{CDF}

\section*{Acknowledgments}

The author thanks G. Kramer, I. Schienbein, and H. Spiesberger for the
collaboration in the work presented here.

\end{document}